\documentclass[preprint,showpacs,preprintnumbers,amsmath,amssymb]{revtex4}

\usepackage{amssymb}

\usepackage{graphicx}
\usepackage{dcolumn}
\usepackage{bm}

\begin{document}

\title{\textbf{Electron Dynamics at a Positive Ion}}
\author{James W. Dufty}
\affiliation{Department of Physics, University of Florida,
Gainesville, FL 32611}
\author{Ilya V. Pogorelov}
\affiliation{Acccelerator and Fusion Research Division, MS 71J,
Lawrence Berkely National Laboratory,
Berkeley, CA 94720}
\author{Bernard Talin and Annette Calisti}
\affiliation{UMR6633, Universit\'{e} de Provence, Centre Saint J\'{e}r\^{o}me,\\
13397 Marseille Cedex 20, France.}
\date{\today }

\begin{abstract}
The dynamics of electrons in the presence of a positive ion is
considered for conditions of weak electron-electron couping but
strong electron-ion coupling. The equilibrium electron density and
electric field time correlation functions are evaluated for
semi-classical conditions using a classical statistical mechanics
with a regularized electron-ion interaction for MD simulation. The
theoretical analysis for the equilibrium state is obtained from
the corresponding nonlinear Vlasov equation. Time correlation
functions for the electrons are determined from the linearized
Vlasov equation. The resulting electron dynamics is described in
terms of a distribution of single electron-ion trajectories
screened by an inhomogeneous electron gas dielectric function. The
results are applied to calculation of the autocorrelation function
for the electron electric field at the ion for $ 0\leq Z\leq 40$,
including conditions of strong electron-ion coupling. The electron
stopping power and self-diffusion coefficient are determined from
these results, and all properties calculated are compared with
those obtained from semi-classical molecular dynamics simulation.
The agreement with semi-classical MD simulation is found to be
reasonable. The theoretical description provides an instructive
interpretation for the strong electron-ion results.
\end{abstract}
\maketitle

\section{Introduction}

The total electric field at a particle in a plasma determines the
dominant radiative and transport properties of that particle. The
theory for the equilibrium distribution of fields at a neutral or
charged point is well developed \cite{Duftymicro}. The theory for
the dynamics of such fields is more complicated
\cite{Zogaib0,Alastuey} and some progress has been made recently
in special cases. For example, the field dynamics at a neutral
point has been described exactly in the Holtzmark limit
\cite{Berkovsky0}. The dynamical properties of fields due to
\emph{positive} ions near a \emph{positive} impurity have been
given an accurate approximate evaluation for a wide range of
charge coupling, relative charge numbers, and relative masses
\cite{Boercker,Berkovsky}. The corresponding study of
\emph{negative} charges (electrons) at a positive ion has been
considered more recently only for the simplest case of a single
ion of charge number $Z$ in a semiclassical electron gas
\cite{Talin,Talin0}. In this case, the attractive interaction
between the electron and ions emphasizes further the nonlinear
dependence on $Z$. The static properties (electron charge density,
electron microfield distribution) have been discussed in some
detail for this case elsewhere \cite{Talin}. Here, attention is
focused on the dynamics via the equilibrium electron electric
field autocorrelation function. This case of electron fields at a
positive ion is qualitatively different from same sign ion fields,
since in the former case electrons are attracted to the ion
leading to strong electron-ion coupling for the enhanced close
configurations.

It is difficult \textit{a priori} to predict even the qualitative
features of the field autocorrelation functions due to this
inherent strong electron-ion coupling, and there is no
phenomenology for guidance. Consequently, our initial analysis has
been based on MD simulation of the correlation functions followed
by an attempt to model and interpret the observed results.
However, MD simulation for the electrons is limited to classical
mechanics, while the singular attractive electron-ion interaction
inherently requires a quantum mechanical description. This
difficulty is circumvented by modifying the electron-ion Coulomb
potential at short distances to represent quantum diffraction
effects. The conditions for validity and limitations of this
classical model have been discussed extensively elsewhere
\cite{Bonitz}. The details of the MD method also have been
discussed elsewhere \cite{Talin} and will not be repeated here.
There is a growing recent literature on the related MD studies of
two-component classical models of a hydrogen plasma
\cite{Zwick0,Norman} at stronger electron-electron coupling
values, but restricted to $Z=1$. Thus the main new feature studied
here is the dependence of structure and dynamics on charge number
$Z.$

The relevant dimensionless parameters are the charge number of the
ion, $Z$, the electron-electron coupling constant $\Gamma
=e^{2}/r_{0}k_{B}T$, and the de Broglie wavelength relative to the
interelectron distance, $\delta =( 2\pi \hbar
^{2}/m_{e}k_{B}Tr_{0}^{2}) ^{1/2}$. The interelectron distance is
defined in terms of the electron density $n_{e}$ by $4\pi
n_{e}r_{0}^{3}/3=1$. The electron-ion coupling is measured by the
maximum value of the magnitude of the regularized electron-ion
potential at the origin, $\sigma =Z\Gamma /\delta $. Most of the
results described below are for $\Gamma =0.1$, $\delta =0.4$, and
$\sigma =0.25Z$, with $0\leq Z\leq 40$. The corresponding density
and temperature are $n_{e}=2.5\times 10^{22}$ cm$^{-3}$ and
$T=7.9\times 10^{5}$ $K$. Additional results are presented for the
(unrealistic) extreme electron-ion coupling conditions of $\Gamma
=$ $0.5$, $\delta =0.2$, and $\sigma =2.5Z$, with $0\leq Z\leq 10$
($n=3.2\times 10^{18}$ cm$^{-3}$, $ T=7.9\times 10^{3}$ $K$ ), and
for the experimentally relevant conditions of $\Gamma =0.029$,
$\delta =0.059$, and $\sigma =0.5Z$, with $0\leq Z\leq 8$ ($
n=1\times 10^{19}$ cm$^{-3}$, $T=2\times 10^{5}$$K$ ). In all
cases the electron-electron coupling is weak. Since $\Gamma $ is
small the kinetic equation for the electron reduced distribution
function becomes the nonlinear Vlasov equation in the presence of
the external ion potential. For the equilibrium state this
equation gives the nonlinear Boltzmann-Poisson equation. It is
mathematically equivalent to the HNC integral equations for an
impurity in a one component plasma \cite{HNC}, applicable as well
for larger $\Gamma $, and can be solved numerically in a similar
way. However, as noted elsewhere \cite{Talin}, these numerical
methods fail at very strong coupling $(\sigma \gtrsim 8)$. For
practical purposes, the stationary solution is modelled as a
nonlinear Debye-Huckel distribution with parameters fit to the HNC
solution when it exists or by comparison with MD simulations
otherwise (a variational method also has been described \cite
{Zogaib} but is not used here). The results for equilibrium
properties are shown to be quite accurate.

The equilibrium time correlation functions are determined from the
linearized Vlasov equation (linear perturbations of the
equilibrium state, but still nonlinear in the electron-ion
interaction). The results are a composition of correlated initial
conditions, fields with single electron trajectories about the
ion, and dynamical collective screening by the inhomogeneous
electron distribution about the ion. These results are simplified
further by a mean field approximation resulting in a single
electron problem in the effective potential of the nonlinear Debye
distribution. This analysis is applied to the electric field
autocorrelation function showing reasonable agreement with the
results from MD simulation. The primary observations from the
simulations of the field autocorrelation function for increasing
charge number are: 1) an increase in the initial value, 2) a
decrease in the correlation time, and 3) an increasing
anticorrelation at longer times. The simple mean field
approximation reproduces all of these results. Furthermore, its
simplicity allows an interpretation from the single particle
motion.

A closely related property is the stopping power for an ion by the
electron gas. In the low velocity limit, and for large ion mass,
the stopping power is proportional to the time integral of the
field autocorrelation function \cite{Berkovsky2}. This integral
also determines the self-diffusion and friction coefficients in
this same limit \cite{Berkovsky3}. Linear response predicts a
dominant $Z^{2}$ dependence for these properties. Significant
deviations from this $Z^{2}$ dependence are observed at strong
coupling and have been the focus of attention in recent years
\cite{Zwick}. The results here show these deviations come from a
competition between the increase of the integral due to 1) above
and the decrease due to 2) and 3). MD simulation shows that the
latter two dynamical effects dominate the former static effect.
The simple mean field model provides the missing interpretation
for 2) and 3).

This same integral of the field autocorrelation function
determines the half width for spectal lines from ion radiators due
to perturbations by electrons in the fast fluctuation limit \cite{Kubo}.
An extension of the analysis provided here to this atomic physics problem
is under way and promises to provide an additional experimental probe
for the dynamics of charges near an ion \cite{Talin2,Talin3}.

The theoretical description based on the Vlasov equation is
provided in the next section. The results are applied to the field
autocorrelation function in section 3. The stopping power and
self-diffusion coefficient are evaluated in section 4. Finally, a
summary and discussion is provided in the last section.

\section{Kinetic theory}

The classical system considered consists of $N_{e}$ electrons with
charge $-e $, an infinitely massive positive ion with charge $Ze$
placed at the origin, and a rigid uniform positive background for
overall charge neutrality contained in a large volume $V$. The
Hamiltonian has the form
\begin{equation}
H=\sum_{\alpha =1}^{N_{e}}\left( \frac{p_{\alpha
}^{2}}{2m}+V_{ei}\left( r_{\alpha }\right) +V_{eb}\left( r_{\alpha
}\right) \right) +\frac{1}{2} \sum_{\alpha ,\gamma }^{N_{e}}
V_{ee}(\left| \mathbf{r}_{\alpha }-\mathbf{r} _{\gamma }\right| )
\label{2.1}
\end{equation}%
where $V_{ee}(\left| \mathbf{r}_{\alpha }-\mathbf{r}_{\gamma
}\right| )$is the Coulomb interaction for electrons $\alpha $ and
$\gamma $, $V_{ei}\left( r_{\alpha }\right) $ is the regularized
electron-ion interaction for electron $\alpha$ with the ion, and
$V_{eb}\left( r_{\alpha}\right) $ is the Coulomb interaction for
electron $\alpha$ with the uniform neutralizing background
\begin{equation}
V_{ee}(\left| \mathbf{r}_{\alpha }-\mathbf{r}_{\gamma }\right|
)=\frac{e^{2}}{\left| \mathbf{r}_{\alpha }-\mathbf{r}_{\gamma
}\right| },\hspace{0.3in} V_{ei}(r_{\alpha })=-Ze^{2}\
\frac{\left( 1-e^{-r_{\alpha }/r_{0}\delta }\right) }{r_{\alpha
}}.  \label{2.2}
\end{equation}
\begin{equation}
V_{eb}(r_{\alpha })=-\left( 1-\frac{Z}{N_{e}}\right) n_{e}\int
d\mathbf{r} V_{ee}(\left| \mathbf{r}_{\alpha}-\mathbf{r}\right| ).
\label{2.3}
\end{equation}
For values of $r/r_{0}>>\delta $ the potential $V_{ei}(r)$ becomes
Coulomb, while for $r<<\delta $ the Coulomb singularity is removed
and $ \beta V^{ei}(r)\rightarrow -\beta
Ze^{2}/r_{0}\delta=-=-Z\Gamma /\delta =-\sigma $. This is the
simplest phenomenological form representing the short range
effects of the uncertainty principle \cite{Minoo}. In principle
there should be a similar regularization of the electron-electron
interaction, but since that interaction is repulsive
configurations with a pair of electrons within a thermal de
Broglie wavelength are rare. For simplicity, therefore, the
electron-electron interaction is taken to be Coulomb. In all of
the following the electron-electron coupling (weak) is measured by
$\Gamma$ while that for the electron-ion coupling (possibly
strong) is measured by $\sigma$.

The distribution function for the electrons is denoted by
$f(\mathbf{r}, \mathbf{v};t)$. It is normalized such that
integration over all velocities and the volume of the system
equals $N_{e}$. It obeys the exact first BBGKY (Born, Bogoliubov,
Green, Kirkwood, Yvon) hierarchy equation \cite{McLennan}
\begin{eqnarray}
&&\left( \partial _{t}+\mathbf{v\cdot \nabla
}_{\mathbf{r}}-m_{e}^{-1}\left( \mathbf{\nabla
}_{\mathbf{r}}\left( V_{ei}(\mathbf{r})+V_{eb}(\mathbf{r} )\right)
\right) \cdot \mathbf{\nabla }_{\mathbf{v}}\right) f(\mathbf{r},
\mathbf{v};t)  \nonumber \\
&=&m_{e}^{-1}\int d\mathbf{r}_{2}d\mathbf{v}_{2}\left(
\mathbf{\nabla }_{ \mathbf{r}}V_{ee}(\mathbf{r-r}_{2})\right)
\cdot \mathbf{\nabla }_{\mathbf{v}
}f^{(2)}(\mathbf{r},\mathbf{v};\mathbf{r}_{2},\mathbf{v}_{2};t)
\label{2.4}
\end{eqnarray}
Here
$f^{(2)}(\mathbf{r},\mathbf{v};\mathbf{r}_{2},\mathbf{v}_{2};t)$
is the joint distribution function for two electrons. At the weak
coupling ($\Gamma <<1$) the electron distributions are
approximately independent and
$f^{(2)}(\mathbf{r},\mathbf{v};\mathbf{r}_{2},\mathbf{v}_{2};t)
\rightarrow$
$f(\mathbf{r},\mathbf{v};t)$$f(\mathbf{r}_{2},\mathbf{v}_{2};t).$
Then (\ref{2.4}) becomes the nonlinear Vlasov equation
\begin{eqnarray}
&&\left( \partial _{t}+\mathbf{v\cdot \nabla }_{\mathbf{r}}
-m_{e}^{-1}\left(\mathbf{\nabla }_{\mathbf{r}}\left(
V_{ei}(\mathbf{r})\right) \right) \cdot
\mathbf{\nabla }_{\mathbf{v}}\right) f(\mathbf{r},\mathbf{v};t)
\nonumber \\
&=&m_{e}^{-1}\left( \mathbf{\nabla
}_{\mathbf{v}}f(\mathbf{r},\mathbf{v} ;t)\right) \cdot
\mathbf{\nabla }_{\mathbf{r}}\int d\mathbf{r}_{2}V_{ee}(
\mathbf{r-r}_{2})\delta n(\mathbf{r}_{2},t)  \label{2.5}
\end{eqnarray}
where $\delta n(\mathbf{r},t)$ is the deviation of the electron
density from the rigid uniform positive background
\begin{equation}
\delta n(\mathbf{r},t)=n(\mathbf{r},t)-\left(
1-\frac{Z}{N_{e}}\right) n_{e},
\hspace{0.25in}n(\mathbf{r},t)=\int
d\mathbf{v}f(\mathbf{r},\mathbf{v};t). \label{2.6}
\end{equation}

The left side of (\ref{2.5}) describes the single electron motion
in the presence of the ion at the origin. Since the charge number
can be large, the electron ion coupling can be large. The right
side of this equation describes the correlation effects due to
interactions among the electrons, which are weak for hot, dense
matter. However, these weak correlations depend nonlinearly on the
distribution function so that the content of this equation and its
solutions can be quite rich and complex.

\subsection{Equilibrium solution}

The equilibrium state is a stationary solution to (\ref{2.5}).
However, it is known from classical equilibrium statistical
mechanics that it must be of the form
\begin{equation}
f_{e}(\mathbf{r},\mathbf{v})=n_{e}(r)\phi (v),\hspace{0.25in}\phi
(v)=\left( \frac{\beta m}{2\pi }\right) ^{3/2}e^{-\beta mv^{2}/2}.
\label{2.7}
\end{equation}
Substitution of this into (\ref{2.5}) gives the nonlinear integral
equation for $n_{e}(r)$
\begin{equation}
\ln \frac{n_{e}(r)}{n_{e}}=-\beta V_{ei}(r)-\beta e^{2}\int
d\mathbf{r} ^{\prime }\frac{\delta n_{e}(r^{\prime })}{\left|
\mathbf{r}-\mathbf{r} ^{\prime }\right| }  \label{2.8}
\end{equation}
where $n_{e}=N_{e}/V$ is the average volume. It is often useful to
write the solution in terms of an effective electron-ion potential
\begin{equation}
\frac{n_{e}(r)}{n_{e}}\equiv e^{-\beta U_{ei}(r)}  \label{2.8a}
\end{equation}
According to (\ref{2.8}) $U_{ei}(r)$ obeys the nonlinear integral
equation
\begin{equation}
U_{ei}(r)=V_{ie}(r)+n_{e}e^{2}\int d\mathbf{r}^{\prime
}\frac{1}{\left| \mathbf{r}-\mathbf{r}^{\prime }\right| }\left(
e^{-\beta U_{ei}(r^{\prime })}-1\right)   \label{2.8b}
\end{equation}
The second term provides the nonlinear strong coupling effects of
the electron-ion interactions. It is worth noting that although
the electron-electron coupling is weak, the strong ion-electron
effects are mediated by the electron-electron interaction of this
second term. The predictions for a free electron gas interacting
with an ion are quite different from those given below.

The numerical solutions to (\ref{2.8}) or (\ref{2.8b}) and its
relationship to the hypernetted chain integral equation at
stronger electron coupling has been discussed in reference
\cite{Talin}. At the weak electron-electron coupling considered
here these equations are the same as the HNC equations, and their
solution in the following will be referred to as the HNC result.
An important observation in reference \cite{Talin} is
that the numerical solution is well represented by the Debye form%
\begin{equation}
U_{ei}(r^{\prime })=\frac{-\overline{Z}e^{2}}{\left( 1-\left(
\delta / \overline{\lambda }\right) ^{2}\right) }\frac{1}{r}\left(
e^{-r/r_{0} \overline{\lambda }}-e^{-r/r_{0}\delta }\right) .
\label{2.9}
\end{equation}
where the effective charge number $\overline{Z}$ and screening
length $ \overline{\lambda }$ are fitting parameters. In the weak
coupling domain these become the actual charge number and the
Debye screening length and this form is exact. More generally, it
is only an approximation and the best choices for $\overline{Z}$,
$\overline{\lambda }$ are different from $Z$, $\lambda $ at strong
coupling. This approximation for $n_{e}(r)$ will be referred to as
the nonlinear Debye approximation. Figure 1 illustrates the
fitting of the nonlinear Debye form to HNC for the case $Z=8$,
$\Gamma =0.1$, and $\delta =0.4$. Also shown are the corresponding
results from MD simulation. The agreement is quite good. Finally,
the linear Debye form is shown to indicate that nonlinear effects
are clearly significant. Similar results are obtained for the
other values of $Z$ discussed below.
\begin{figure}[t]
\includegraphics[width=0.8\columnwidth]{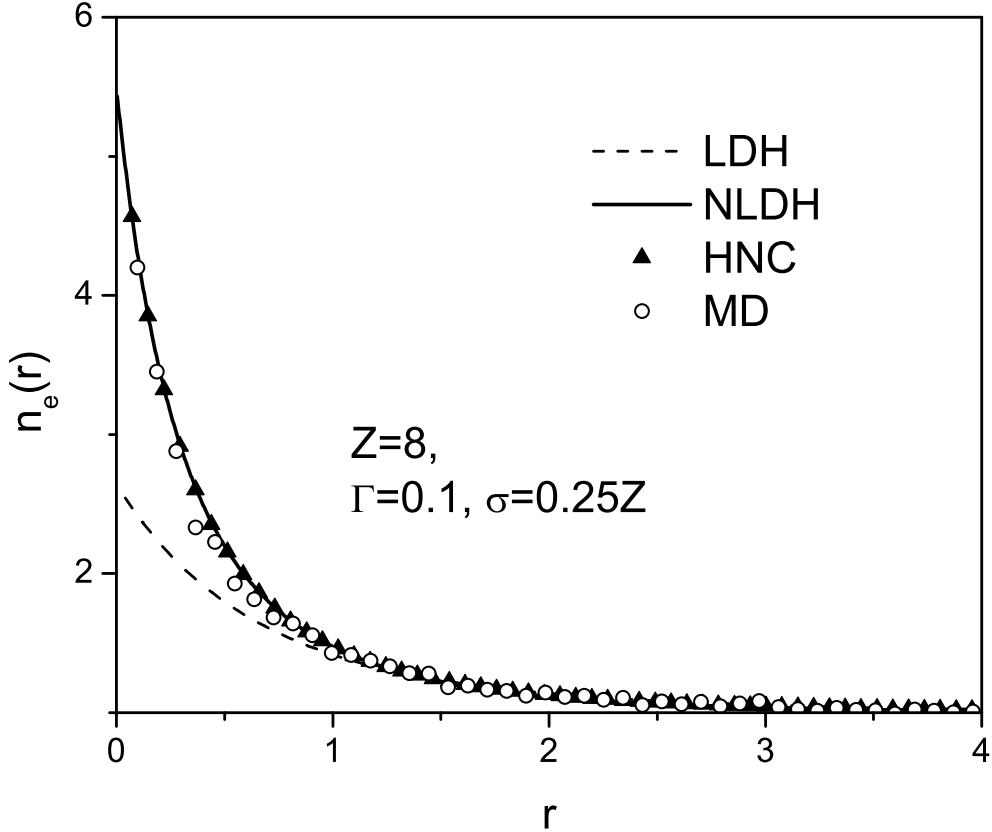}
\caption{Electron density around an ion of charge number $Z=8$}.
\end{figure}

\subsection{Electric field time correlation function}

To explore the dynamics of the electrons attention is limited to
the electric field autocorrelation function defined by
\begin{equation}
C(t)=\frac{r_{0}^{4}}{e^{2}}<\mathbf{E}\left( t\right)
\mathbf{\cdot E}>. \label{2.10}
\end{equation}
The brackets denote an equilibrium Gibbs ensemble average and
$\mathbf{E}$ is the total field at the ion due to all electrons
\begin{equation}
\mathbf{E}=\frac{1}{Ze}\sum_{\alpha =1}^{N_{e}}\nabla _{r_{\alpha
}}V_{ei}\left( r_{\alpha }\right) =\sum_{\alpha
=1}^{N_{e}}\mathbf{e}\left( \mathbf{r}_{\alpha }\right)
\label{2.11}
\end{equation}
\begin{equation}
\mathbf{e}\left( \mathbf{r}_{\alpha }\right)
=e\frac{\widehat{\mathbf{r}} _{\alpha }}{r_{\alpha }^{2}}(1-(
1+\frac{r_{\alpha }}{r_{0}\delta }) e^{-r_{\alpha }/r_{0}\delta })
\label{2.12}
\end{equation}
Consider first the initial value $C(0)$ for which two exact
representations can be given
\begin{eqnarray}
C(0) &=&\frac{r_{0}^{4}}{e^{2}}\int d\mathbf{re}\left(
\mathbf{r}\right) \mathbf{\cdot }\left[ n_{e}(r)\mathbf{e}\left(
\mathbf{r}\right) +\int d\mathbf{r}^{\prime
}n_{e}(\mathbf{r},\mathbf{r}^{\prime })\mathbf{e}\left(
\mathbf{r}^{\prime }\right) \right]   \nonumber \\
&=&\frac{r_{0}^{4}}{e^{2}}\int d\mathbf{r}n_{e}(r)\mathbf{e}\left(
\mathbf{r} \right) \mathbf{\cdot e}_{mf}\left( \mathbf{r}\right)
\label{2.12a}
\end{eqnarray}
The first equality expresses the covariance in terms of both the
electron density $n_{e}(r)$ and the density for two electrons near
the ion $n_{e}(\mathbf{r},\mathbf{r}^{\prime })$. The second
representation requires only the electron density and the mean
force field derived from it
\begin{eqnarray}
\mathbf{e}_{mf}\left( \mathbf{r}\right)
&=&\frac{1}{n_{e}(r)}\left[n_{e}(r)
\mathbf{e}\left(\mathbf{r}\right) +\int d\mathbf{r}^{\prime
}n_{e}(\mathbf{r},\mathbf{r}^{\prime})\mathbf{e}\left(\mathbf{r}^{\prime
}\right) \right]
\nonumber \\
&=&\frac{1}{\beta Ze}\nabla \ln \frac{n_{e}(r)}{n_{e}}.  \label{2.14}
\end{eqnarray}
The second equality is derived in Appendix A. It is similar to the
first form with the \emph{apparent} neglect of the two-electron
ion correlations. However, these latter contributions are
incorporated exactly in the mean force field
$\mathbf{e}_{mf}\left(\mathbf{r}\right)$. The effects of
electron-electron interactions are included in $n_{e}(r)$ through
the mean field screening of the ion-electron interaction. However,
$n_{e}(\mathbf{r},\mathbf{r}^{\prime })$ describes the additional
electron correlations for two electrons near the ion beyond these
mean field effects.
\begin{figure}[t]
\includegraphics[width=0.8\columnwidth]{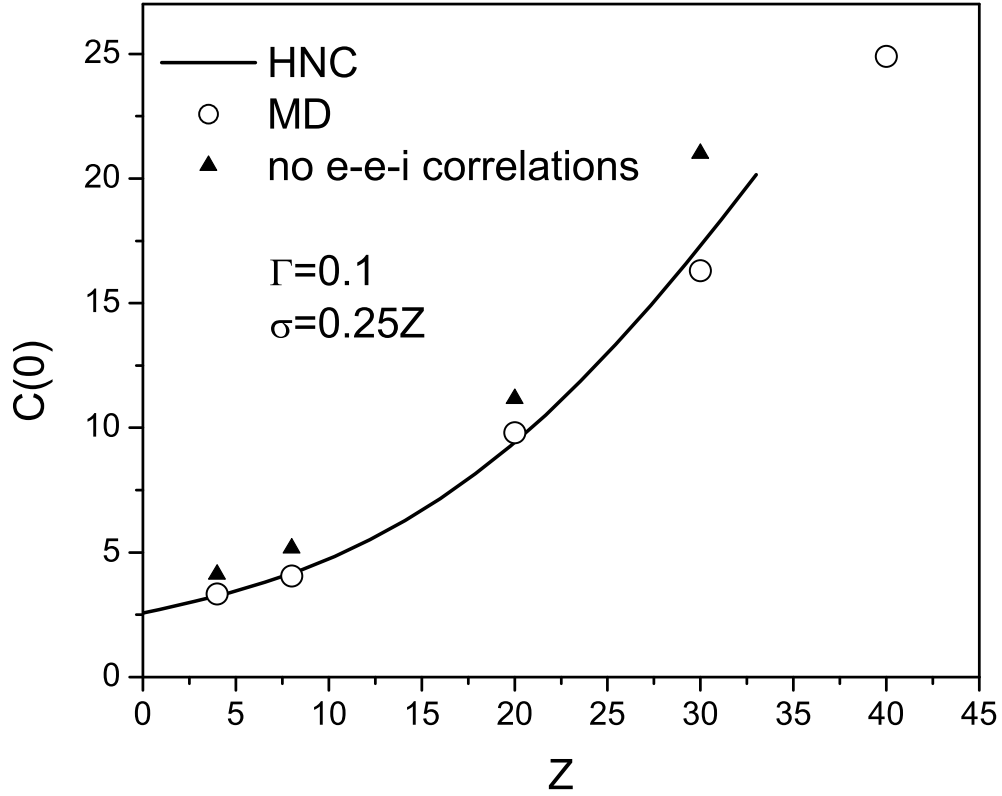}
\caption{Initial value for the electric field autocorrelation
function}.
\end{figure}
Figure 2 shows the initial field fluctuations $C(0)$ as a function
of $Z$, at $\Gamma =0.1$ and $\sigma =0.25Z$, calculated using
$n_{e}(r)$ from HNC and calculated directly from MD. The solid
curve is the HNC data indicating a strong $\sim Z^{3}$ dependence
inherited from the electron charge density. Also shown are the
results from the first equality of (\ref{2.12a}) neglecting the
correlations of two electrons in the presence of the ion,
$n_{e}(\mathbf{r},\mathbf{r}^{\prime}) \rightarrow
n_{e}(\mathbf{r})n_{e}(\mathbf{r}^{\prime})/n_e$. Clearly, these
are significant for the conditions considered, showing the
coupling between strong ion-electron interactions and weak
electron-electron interactions. The results are similar for the
other coupling conditions considered ($\sigma =0.5Z$ and $\sigma
=2.5Z$).

It is shown in Appendix B that the time dependence of $C(t)$ in
the weak coupling limit can be written as
\begin{equation}
C(t)=\int d\mathbf{r}d\mathbf{ve}(\mathbf{r})\cdot \mathbf{\psi
}(\mathbf{r}, \mathbf{v},t)  \label{2.15}
\end{equation}
where $\mathbf{\psi }(\mathbf{r},\mathbf{v},t)$ obeys the
linearized Vlasov equation
\begin{equation}
\left( \partial _{t}+\mathcal{L}\right) \mathbf{\psi
}(\mathbf{r},\mathbf{v} ;t)=-\beta
f_{e}(\mathbf{r},\mathbf{v})\mathbf{v}\cdot \mathbf{\nabla }_{
\mathbf{r}}\int d\mathbf{r}_{2}V_{ee}(\mathbf{r-r}_{2})\int
d\mathbf{v}_{2} \mathbf{\psi }(\mathbf{r}_{2},\mathbf{v}_{2},t).
\label{2.16}
\end{equation}
The associated initial condition is
\begin{equation}
\mathbf{\psi
}(\mathbf{r},\mathbf{v},t=0)=\mathbf{e}_{mf}(\mathbf{r})f_{e}(
\mathbf{r},\mathbf{v}),  \label{2.17}
\end{equation}
where $\mathbf{e}_{mf}(\mathbf{r})$ is the mean field of
(\ref{2.14}) above. The linear operator $\mathcal{L}$ is the
generator for single electron dynamics in the effective potential
$U_{ei}(\mathbf{r})$
\begin{equation}
\mathcal{L}=\mathbf{v\cdot \nabla }_{\mathbf{r}}-m_{e}^{-1}\left(
\mathbf{ \nabla }_{\mathbf{r}}\left( U_{ei}(\mathbf{r})\right)
\right) \cdot \mathbf{ \nabla }_{\mathbf{v}}.  \label{2.18}
\end{equation}
The equilibrium distribution is stationary under this operator,
$\mathcal{L} f_{e}(\mathbf{r},\mathbf{v})=0$. In addition to this
effective single particle dynamics, all dynamical many-electron
effects are contained in the term on the right side of
(\ref{2.16}). The details of the formal solution to this kinetic
equation for the correlation function are given in Appendix B with
the result
\begin{equation}
C(t)=\int_{0}^{t}dt^{\prime }\int
d\mathbf{r}d\mathbf{v}f_{e}(\mathbf{r},
\mathbf{v})\mathbf{e}_{s}(\mathbf{r};t^{\prime
})\mathbf{e}_{mf}(\mathbf{r} (t-t^{\prime }))  \label{2.19}
\end{equation}
The field $\mathbf{e}(\mathbf{r}(t-t^{\prime }))$ is given by
(\ref{2.12}) with the initial position shifted to
$\mathbf{r}(t-t^{\prime })$ according to the effective single
particle dynamics generated by $\mathcal{L},$ using the initial
conditions $\mathbf{r},\mathbf{v}$. The other field $\mathbf{e}
_{s}(\mathbf{r};t^{\prime })$ is a dynamically screened field
\begin{equation}
\mathbf{e}_{s}(\mathbf{r};t)=(2\pi )^{-3}\int
d\mathbf{k}e^{-i\mathbf{k} \cdot
r}\widetilde{\mathbf{e}}_{s}(\mathbf{k},t),\hspace{0.25in}\widetilde{
\mathbf{e}}_{s}(\mathbf{k,}t)=\int d\mathbf{k}^{\prime
}\widetilde{\mathbf{e} }(\mathbf{k}^{\prime })\epsilon ^{-1}\left(
\mathbf{k}^{\prime },\mathbf{k},t\right)   \label{2.20}
\end{equation}%
where $\widetilde{\mathbf{e}}(\mathbf{k}^{\prime })$ is the
Fourier transform of (\ref{2.12}), and the dielectric function is
defined by
\begin{equation} \epsilon (\mathbf{k},\mathbf{k}^{\prime },t) =
(2\pi )^{3}\delta (\mathbf{k}-\mathbf{k}^{\prime }) +\pi
(\mathbf{k,k} ^{\prime },t) \widetilde{V}_{ee}( k^{\prime })
\label{2.21}
\end{equation}

\begin{equation}
\pi (\mathbf{k,k}^{\prime },t) =\beta \frac{d}{dt}\int d\mathbf{r
}d\mathbf{v}f_{e}(\mathbf{r},\mathbf{v})e^{i\mathbf{k\cdot
r}}e^{-i\mathbf{k}^{\prime }\mathbf{\cdot r(}t\mathbf{)}}
\label{2.22}
\end{equation}
For $Z=0$ this is the familiar classical random phase
approximation for a uniform electron gas, diagonal in
$\mathbf{k},\mathbf{k}^{\prime }$. More generally, the $Z$
dependence leads to a nonuniform electron density near the ion and
the polarization function $\pi (\mathbf{k,k}^{\prime },t) $
depends on the details of this distribution. It vanishes at $t=0 $
indicating no initial screening, $\pi (\mathbf{k,k}^{\prime },t=0)
=0$. At later times the polarization function is non-zero giving a
space and time dependent additional screening. Further
simplification of this result for practical evaluation is
discussed below.

\section{Molecular dynamics simulation}

The application of standard MD simulation methods using the
semi-classical electron-ion potential is somewhat more complex
than for the usual classical fluids with short range repulsive
interactions. Although finite at short distances the attractive
electron-ion potential allows bound and metastable states for
electrons orbiting round the ion over extended periods. For most
properties, e.g., structural properties, this is not a severe
problem except at low temperatures. However, the interest here is
in electric field dynamics which is very sensitive to close
electron-ion configurations. The protocol for control of anomalous
states with quasi-bound dynamical states has been described in
reference \cite{Talin} and will not be repeated here.
\begin{figure}[t]
\includegraphics[width=0.8\columnwidth]{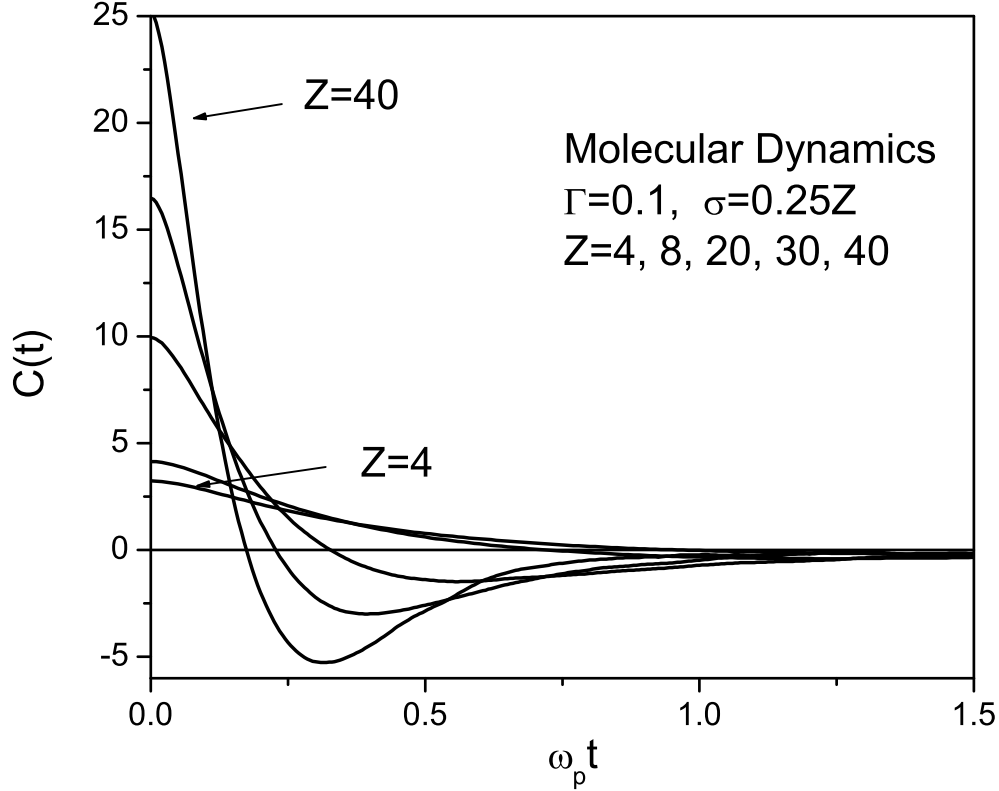}
\caption{Field autocorrelation function at strong coupling}.
\end{figure}
Consider first the coupling conditions $\Gamma =0.1$ and $\sigma
=0.25Z$. Figure 3 shows the field correlation function $C(t)$ for
$Z=4,8,20,30,40$. There are two qualitative features to note. The
first is a characteristic time for relaxation that decreases with
increasing $Z$, and the second is the development of an
anti-correlation that increases with increasing $Z$. Figures 4 and
5 show the corresponding results for the weaker ( $\Gamma =0.029$
and $\sigma =0.5Z$, for\ $Z=1,3,5,8$) and stronger ( $\Gamma =0.5$
and $\sigma =2.5Z$, for\ $Z=1,2,3,4,6,10$) coupling cases. In the
former case the decreasing relaxation time is evident but the
anti-correlation is significant only for the $Z=8$ curve. The
strongest coupling case of Fig. 5 shows large anti-correlation for
all $Z$. It should be noted that this last case is unrealistic
since the equilibrium population of ions at such strong coupling
is extremely small.
\begin{figure}[t]
\includegraphics[width=0.8\columnwidth]{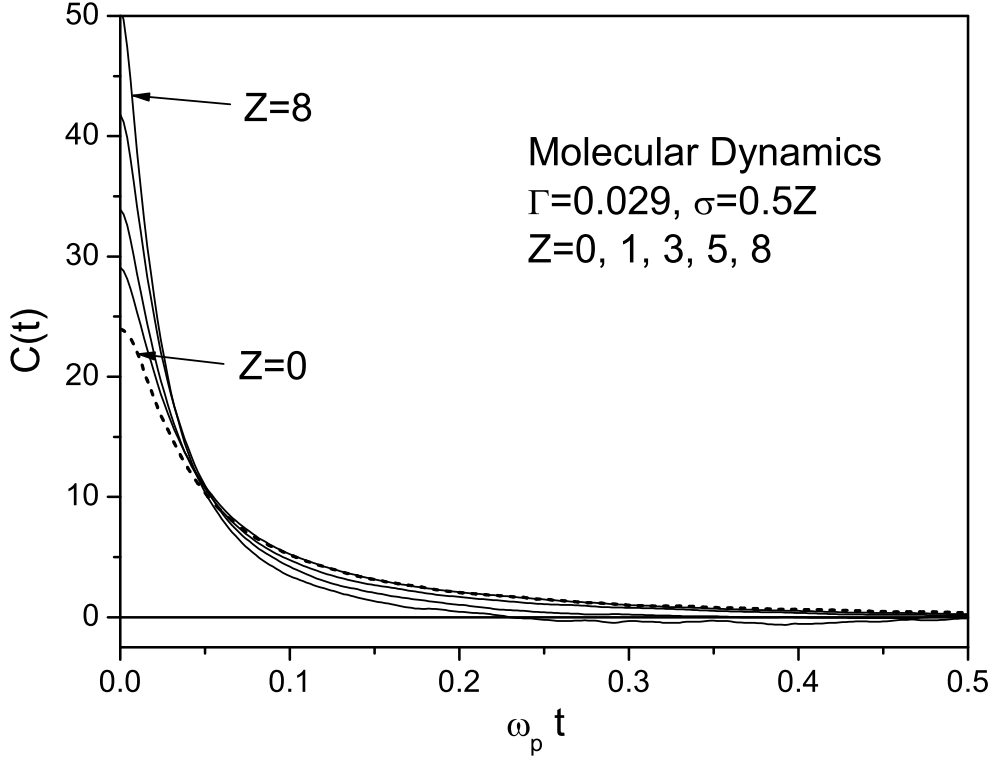}
\caption{Field autocorrelation function at moderate coupling}.
\end{figure}
\begin{figure}[t]
\includegraphics[width=0.8\columnwidth]{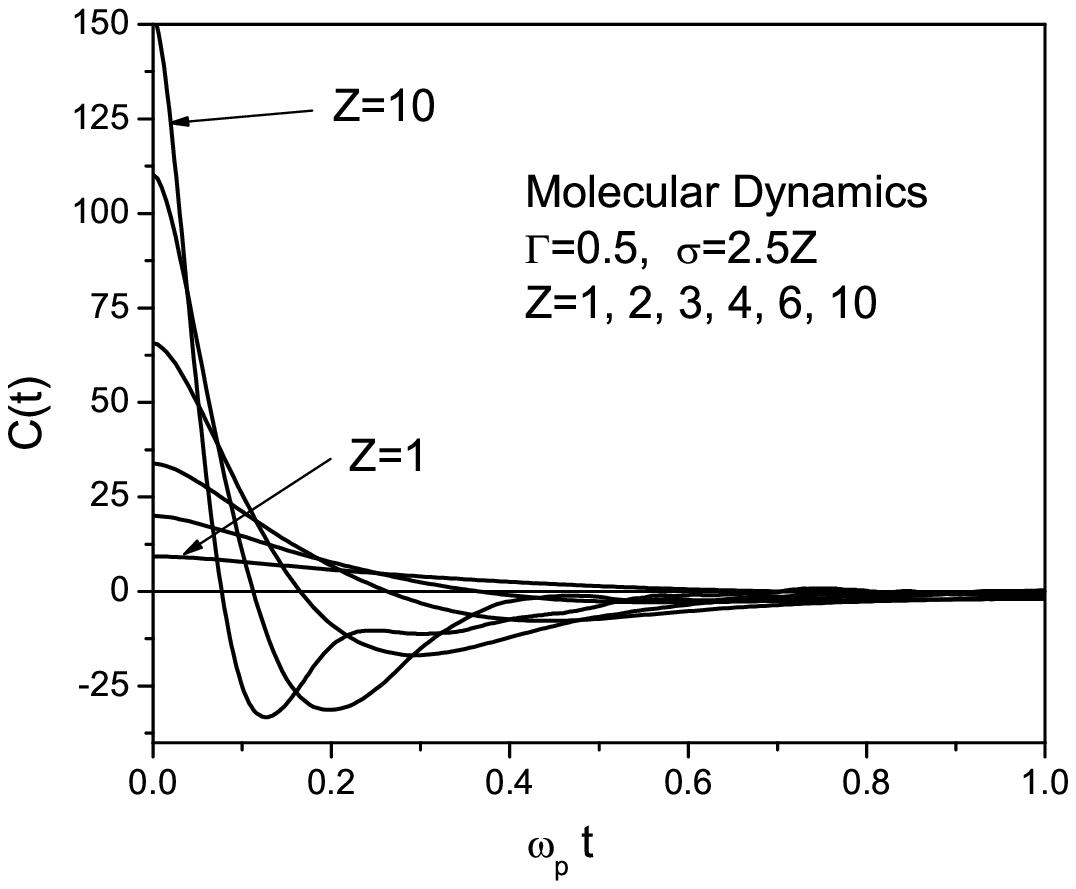}
\caption{Field autocorreltion function at very strong coupling}.
\end{figure}
\begin{figure}[t]
\includegraphics[width=0.8\columnwidth]{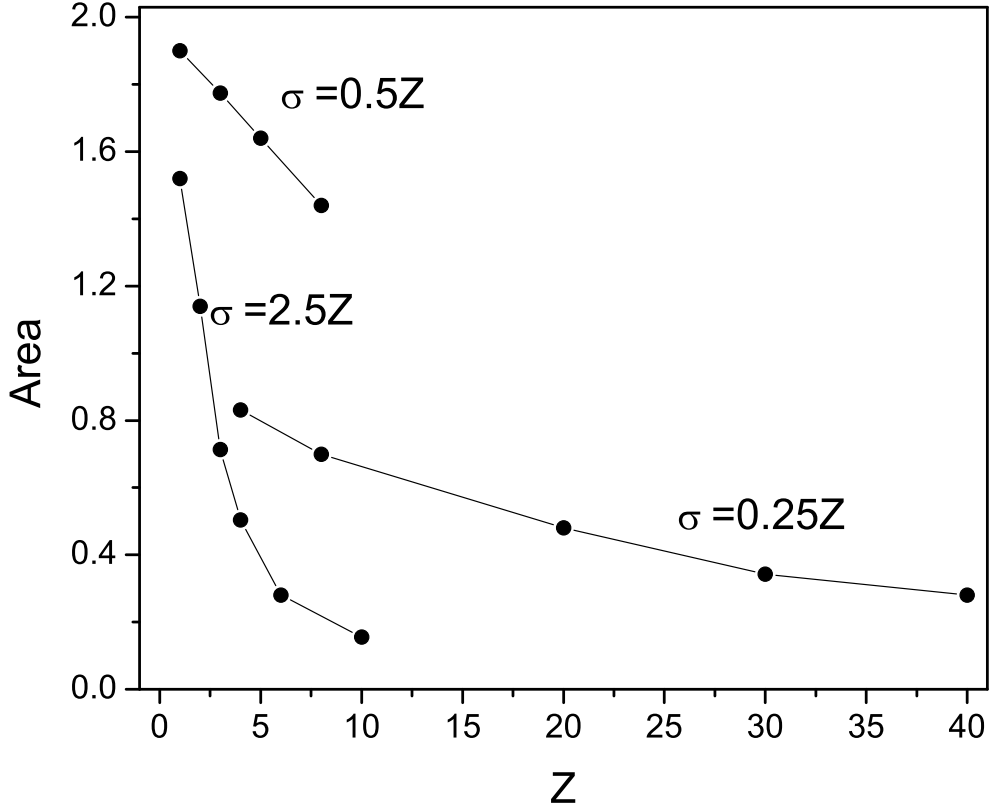}
\caption{Time integral of $C(t)$ as a function of $Z$ for
moderate, strong, and very strong coupling cases}.
\end{figure}
\begin{figure}[t]
\includegraphics[width=0.8\columnwidth]{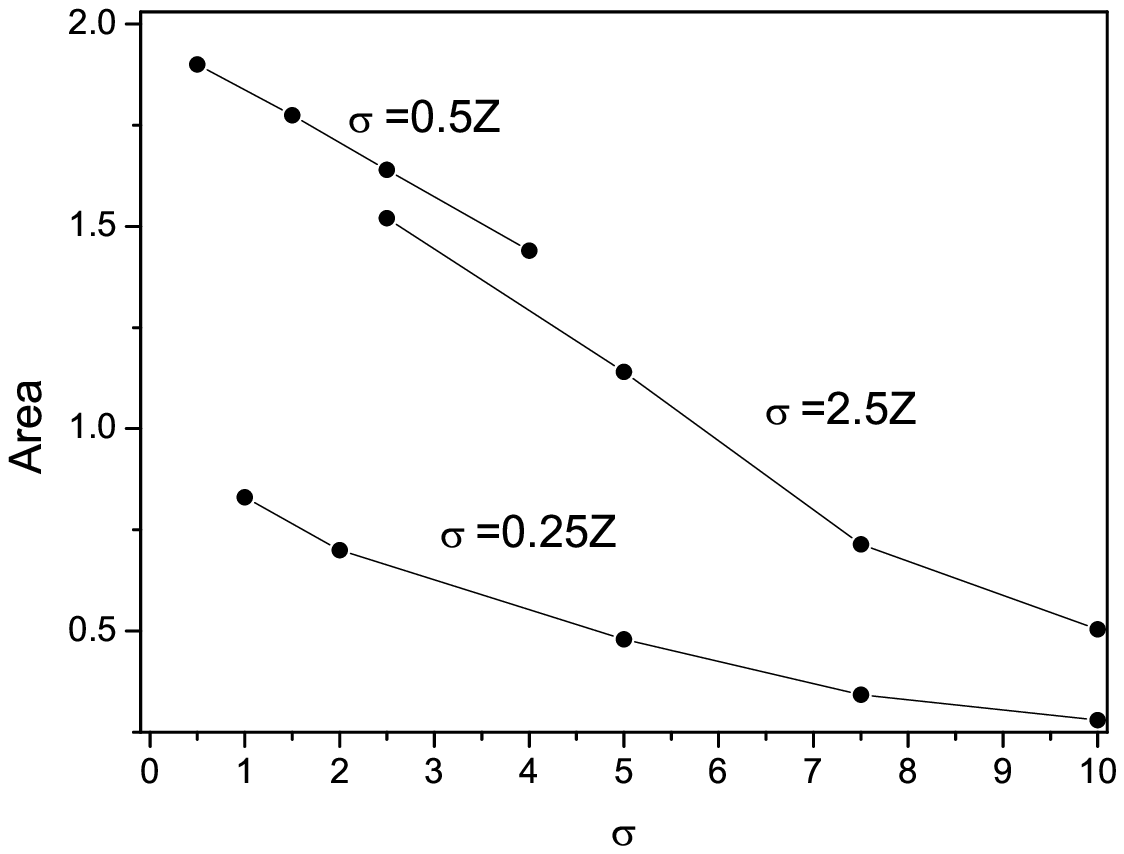}
\caption{Same as Fig. 6 as a function of $\sigma$}.
\end{figure}

The area under the curves is related to transport properties, as
indicated in the next section, and results from a competition
between these two features and the increasing initial
correlations. The effects of shortening decay time and
anti-correlation dominate to decrease the area as indicated in
Fig. 6. These effects are greater for stronger coupling, with
significant anti-correlation occurring for $Z \geq 4$. Figure 7
shows the same results plotted as a function of $\sigma$. The
three cases appear similar, with only a shift in their $\sigma=0$
value due to different values for $\gamma$ and $\delta$. This
suggests that the time integral of the correlation function,
normalized to its value at $\sigma=0$ may be a "universal"
function of sigma.

\section{Effective single particle model}

To calculate the correlation function and identify the mechanisms
for the decay time and correlation, it might be supposed that the
single nearest electron dominates since its field is greatest.
Figure 8 shows this nearest neighbor contribution from MD for the
same conditions of Fig. 3.
\begin{figure}[t]
\includegraphics[width=0.8\columnwidth]{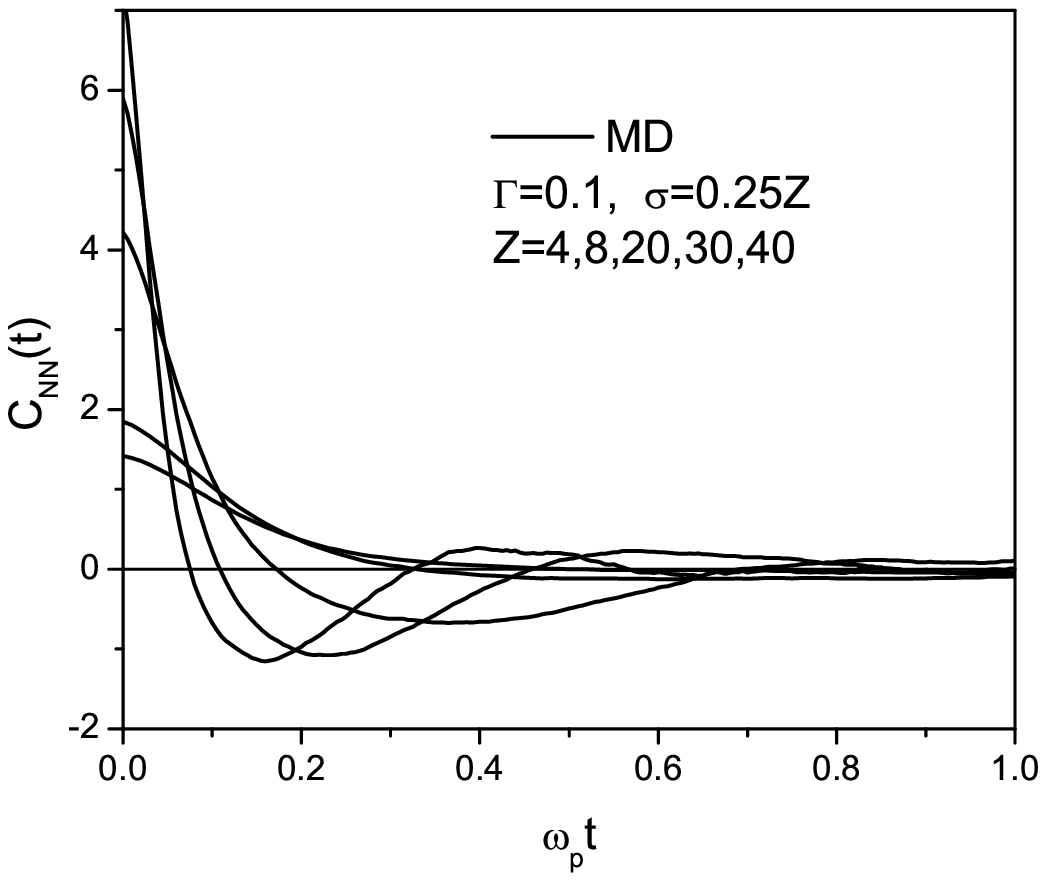}
\caption{Nearest neighbor field autocorrelation function for strong
coupling}.
\end{figure}
Clearly, the nearest neighbor field autocorrelation function has
the same qualitative behavior as the total correlation function
with respect to the decreasing decay time and increasing
anti-correlation. The quantitative values are wrong (both
amplitude and time scale), however, suggesting that contributions
from other particles are important as well. To explore this in
more detail, the Vlasov equation given by (\ref{2.19}) can be
used. It requires evaluation of the dynamically screened field
$\mathbf{e}_{s}( \mathbf{r};t)$, although at short times
$\mathbf{e}_{s}(\mathbf{r};t)\sim
\mathbf{e}(\mathbf{r};0)=\mathbf{e}(\mathbf{r})$. The correlation
function is then effectively that for a single particle moving in
the self-consistent potential (\ref{2.8b}), averaged over the
initial equilibrium distribution of electrons about the ion. Thus
it is similar to the nearest neighbor approximation but extends it
to include all electrons, including the correct initial
correlations and the correlations of the mean field for the
dynamics. If the additional \emph{dynamical} screening is
neglected for all relevant times, i.e.
\begin{equation}
\mathbf{e}_{s}(\mathbf{r};t)\rightarrow \delta (t)\mathbf{e}(\mathbf{r})
\label{4.1}
\end{equation}
then (\ref{2.19}) becomes
\begin{equation}
C(t)\rightarrow \int
d\mathbf{r}d\mathbf{v}f_{e}(\mathbf{r},\mathbf{v})
\mathbf{e}(\mathbf{r})\mathbf{e}_{mf}(\mathbf{r}(t)).
\label{4.2}
\end{equation}
Comparison with (\ref{2.12a}) shows that this approximation is
exact for $ C(0)$. For practical purposed, this approximation for
$C(t)$ has been evaluated using the analytic nonlinear Debye form
(\ref{2.9}) fitted to the HNC results in both the dynamics and the
electron density. The results for this effective single particle
model are shown in Figs. 9 and 10 for the case $\Gamma =0.1$ and
$\sigma =0.25Z$.
\begin{figure}[t]
\includegraphics[width=0.8\columnwidth]{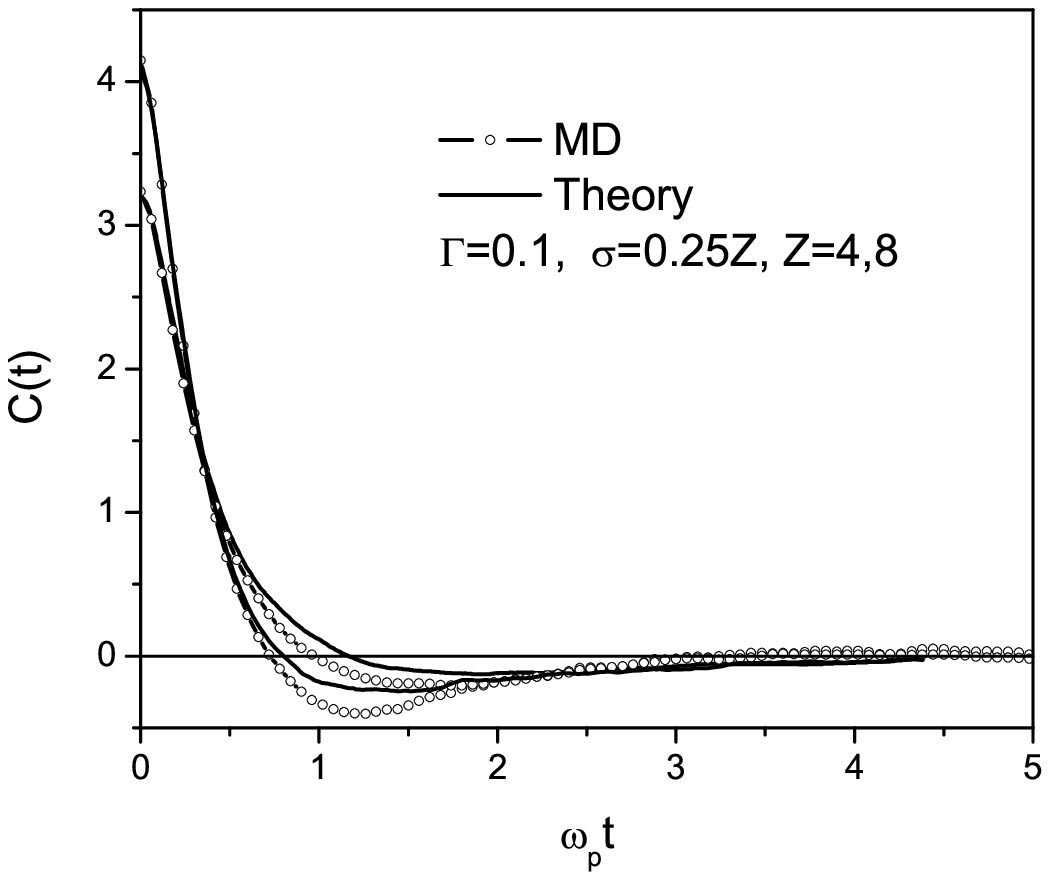}
\caption{Comparison of $C(t)$ from MD with results from mean field
kinetic theory for $Z=4,8$ }.
\end{figure}
\begin{figure}[t]
\includegraphics[width=0.8\columnwidth]{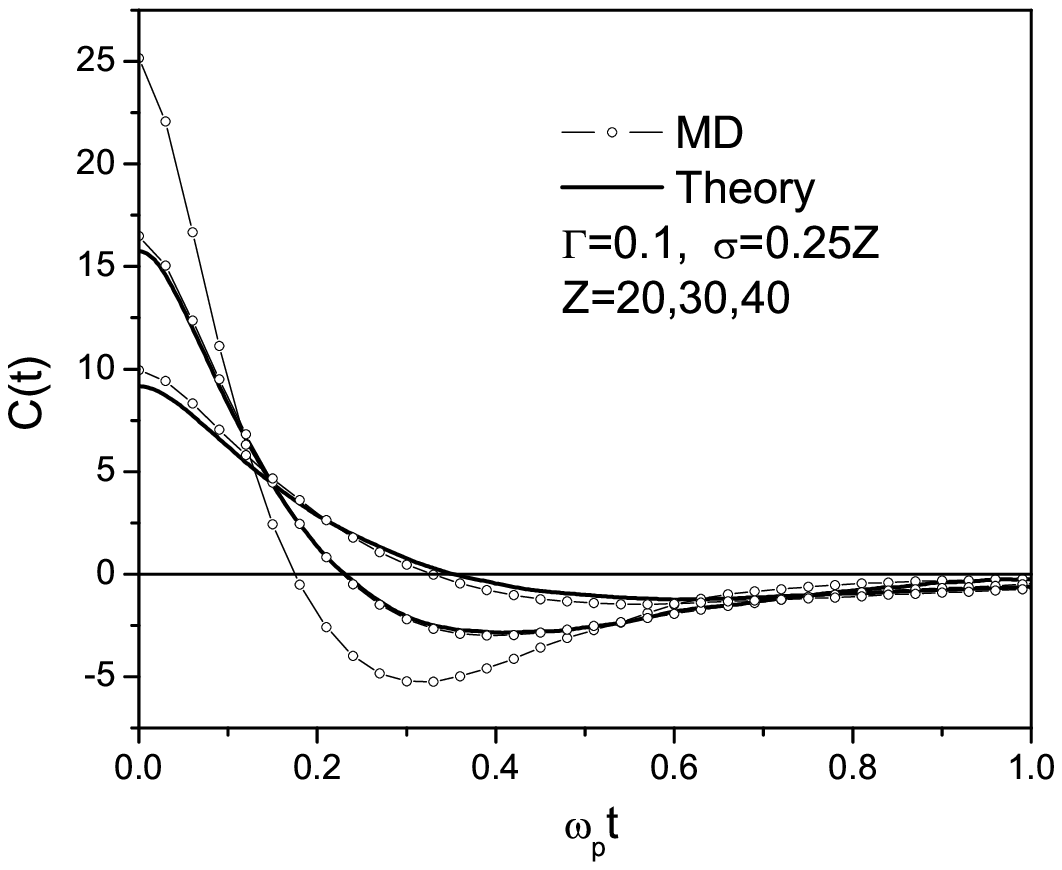}
\caption{Same as Figure 9 for $Z=20,30$}.
\end{figure}

The agreement between this simple theory is quite good and
provides a means to interpret the results of the MD simulation.
The initial position and velocity of an electron are sampled from
the equilibrium distribution $f_{e}(\mathbf{r},\mathbf{v})$ which
favors electrons close to the ion and hence large fields. Since
the force on the electron is also large, its initial acceleration
will be large. This is the source of the short decay time.
Consider the case of an energetic electron near the ion. If its
velocity is directed away from the ion it will move to larger
distances and the field will decrease with a positive component
along the initial field. In contrast, if its initial velocity is
toward the ion it will move past the ion with a change in the
direction of its field relative to the initial value. This is a
source for anti-correlation. For less energetic electrons, the
trajectories are bound and there is continual correlation and
anti-correlation as the correlation function decays in magnitude
due to phase averaging. Both the increase in initial correlation
and the field reversal effects should increase as the charge on
the ion increases, and this is what is observed. At weaker
electron-ion coupling both effects are diminished and blurred as
the relevant configurations are more distant, the fields are
weaker, and the accelerations smaller. This is already evident in
the nearest neighbor results of Fig. 8.

The simple model of (\ref{4.2}) appears better at larger values of
$Z$ where the single particle motion is expected to dominate. For
smaller values of $Z$ the agreement at short times is still good
(the discrepancy at $t=0$ is a limitation of HNC, not the
dynamics), but more significant differences occur after the first
initial decrease. Presumably, this is due to the dynamical
screening effects in $\mathbf{e}_{s}(\mathbf{r};t)$ that have been
neglected. It should be noted that the results here are somewhat
sensitive to the choice of parameters $\overline{Z}$,
$\overline{\lambda }$ used in fitting the non-linear Debye Huckel
form for $f_{e}(\mathbf{r},\mathbf{v})$. Fits emphasizing short or
intermediate distances change slightly the point at which
anti-correlation sets in and its amplitude. The primary criterion
used here was a globally good visual fit and a good resulting
value for the initial condition $C(0)$.

\section{Stopping power, friction, and self-diffusion}

Emphasis here has been placed on the electric field
autocorrelation function as a sensitive measure of electron
properties near the ion. This function is also of interest because
of its connection to transport and radiative properties of the
ion. Specifically, for the case of an infinitely massive ion
considered here there are exact relationships between transport
coefficients characterizing three physically different phenomena:
1) the low velocity stopping power $\mathcal{S}$ for a particle
injected in the electron gas, 2) the friction coefficient $\xi $
for the resistance to a particle being pulled through the gas, and
3) the self-diffusion coefficient $D$ of a particle at equilibrium
with the gas \cite{Berkovsky3}
\begin{equation}
m_{0}\xi =\left( \beta D\right) ^{-1}=\frac{\mathcal{S}(v)}{v}\mid
_{v=0}=\beta Z^{2}r_{0}^{-4}\int_{0}^{\infty }dtC(t)  \label{5.1}
\end{equation}
Finally, the time integral of $C(t)$ also provides the fast
fluctuation limit (impact) for the spectral line width of ions
broadened by electrons \cite{lineshape}. Clearly, a better
understanding of the mechanisms controlling the electric field
autocorrelation function is of interest in several different
contexts.

This Green-Kubo representation (\ref{5.1}) allows a determination
of these transport properties from an equilibrium MD simulation,
as described above. In contrast, previous simulations of stopping
power have studied the nonequilibrium state of the injected
particle, measuring directly the energy degradation \cite{Zwick}.
At asymptotically weak coupling, these properties have a dominant
$Z^{2}$ dependence, as $C(t)$ becomes independent of $Z$. A
puzzling result of the previous simulations \cite{Zwick}, and some
experiments \cite{Experiment}, was the observation of a weaker $Z$
dependence at strong coupling. This behavior is somewhat puzzling
in light of the strong growth of the initial value $C(0)\approx
Z^{3}$ at large $Z$ (see Fig. 2). Thus it would appear that the
dominant dependence would be an even stronger $Z^{5}$. However, as
Fig. 6 shows clearly the competing effects of decreasing
correlation time and a developing time interval of
anti-correlation dominate at strong coupling to decrease the time
integral.
\begin{figure}[t]
\includegraphics[width=0.8\columnwidth]{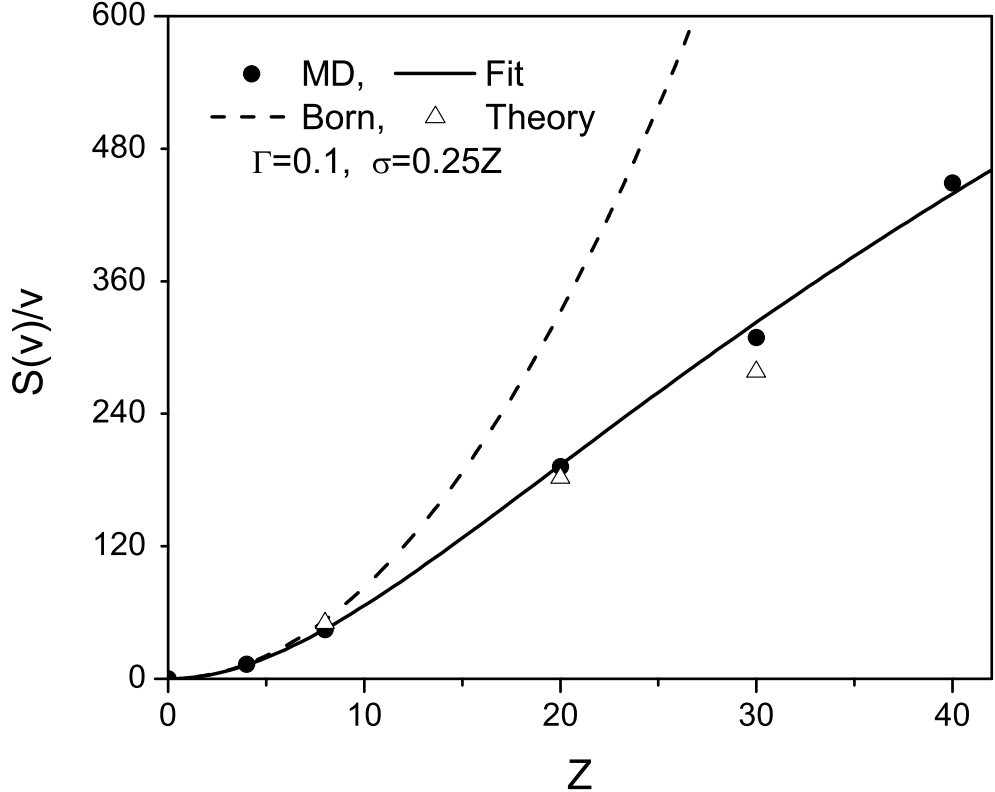}
\caption{Stopping power relative to the speed at zero speed, Eq.
(\ref{5.1}), at strong coupling. Also shown is the Born
approximation and the results of the kinetic theory}.
\end{figure}
Figure 11 shows the dimensionless stopping power as a function of
$Z$ for the case $\Gamma =0.1$ and $\sigma =0.25Z$. Also shown is
the Born approximation $0.83Z^{2}$, where the coefficient has been
determined by the data for small $Z$. The MD data has been fit to
a crossover function
\begin{equation}
\frac{\mathcal{S}(v)}{v}\mid _{v=0}\rightarrow \frac{0.83Z^{2}}{
1+0.008Z^{1/2}}.  \label{5.2}
\end{equation}
This form has been chosen since it implies the stopping power goes
as $Z^{3/2}$ at extreme coupling, which is consistent with the
earlier results \cite{Zwick,Experiment}. However, other fits to
the data here are possible as well. The predictions of the simple
effective single particle theory are shown on Fig. 11 also.
\begin{figure}[t]
\includegraphics[width=0.8\columnwidth]{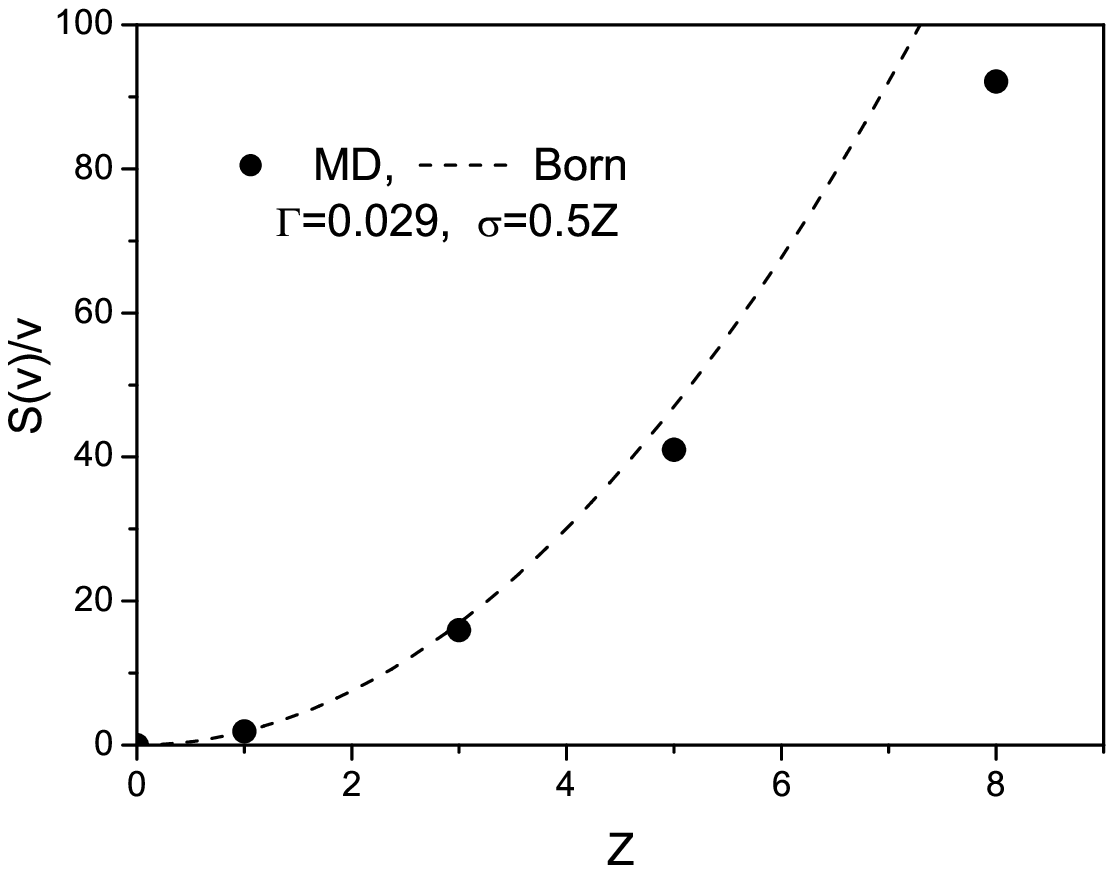}
\caption{Same as Fig. 11 at moderate coupling}.
\end{figure}

\begin{figure}[t]
\includegraphics[width=0.8\columnwidth]{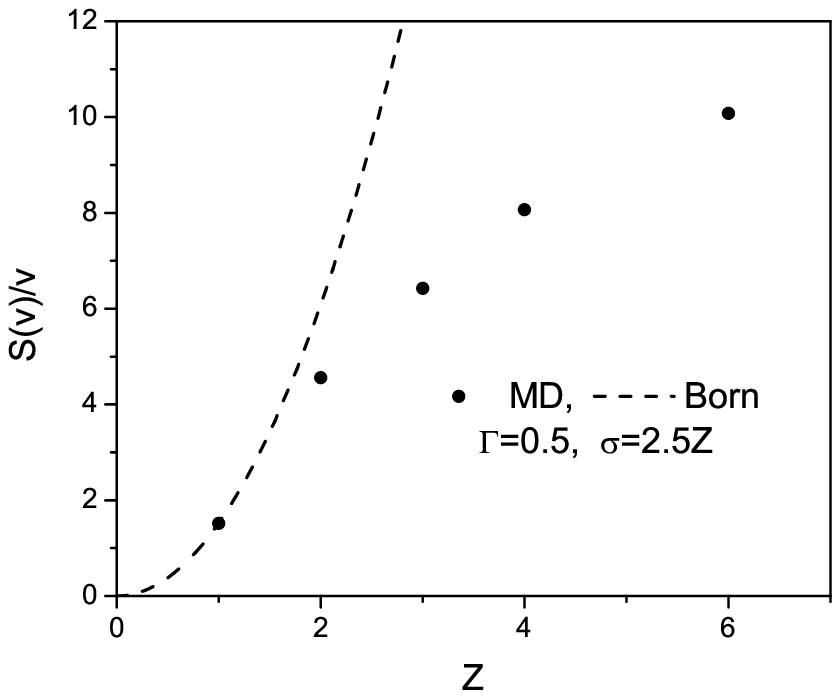}
\caption{Same as Fig. 11 at very strong coupling}.
\end{figure}

Similar results are obtained for the other coupling cases. Figure
12 shows the stopping power for the weaker coupling case of
$\Gamma =0.029$ and $ \sigma =0.5Z$ at smaller values of $Z$. The
Born approximation $1.88Z^{2}$ is determined by the $Z=1$ value.
As expected the Born approximation is quite good at weak coupling,
although some deviation is seen at $Z=8$. This is consistent with
the above estimate that strong coupling effects in Fig. 6 occur
for $\sigma \geq 4.$ The strongest coupling case of $\Gamma =0.5$
and $\sigma =2.5Z$ is shown in Fig. 13, where the Born
approximation is again determined from the $Z=1$ data and found to
be $1.52Z^{2}$. The deviations from the Born approximation are
much greater now, as expected.

\section{Summary and Discussion}

The objective here has been to explore the dynamics of electrons
near a positive ion as a function of the charge number on the ion,
or more precisely, as a function of the electron-ion coupling
$\sigma =Z\Gamma /\delta $. A primary tool for this investigation
has been molecular dynamics simulation, requiring a semi-classical
regularization of the Coulomb potential at short distances. Under
the hot, dense conditions needed to support large $Z$ ions the
electron coupling can be quite weak. An accurate theoretical
description is then given by the nonlinear Vlasov kinetic equation
for the electrons. The MD simulation reveals an interesting
structure for the electric field autocorrelation function. At the
weakest coupling considered there is a rapid initial decay
followed by an asymptotic domain of weak anticorrelation. The time
scale for the initial decay decreases with increasing $Z$ but does
not depend strongly on $Z$ as long as the coupling stays weak.
This is illustrated in Fig. 4. Figures 5 and 6 show a quite
different behavior at strong coupling where the initial decay time
shortens and the anticorrelation becomes prominent on the same
time scale. This qualitative behavior is characteristic of the
single electron dynamics of the nearest neighbor. A quantitative
description is provided by the Vlasov kinetic theory with the
exact initial condition for the distribution of electrons about
the ion, including strong electron-ion correlations mediated by
weak electron-electron interactions. The mean field dynamics is
that of effective single electron trajectories, calculated for the
same effective potential as that for the initial correlations, and
averaged over an ensemble of these initial states.

The simple theoretical model provides a means to interpret the MD
simulation data for the beginnings of a phenomenological
understanding of electron dynamics near a positive ion. The
initial correlation for the electric field $C(0)$ increases with
$Z$ as the equilibrium distribution of electrons is enhanced near
the ion with configurations corresponding to larger fields. The
latter is well described by the nonlinear Debye-Huckel form given
by (\ref{2.8a}) with (\ref{2.9}). The initial decay of $C(t)$ is
essentially the decorrelation time for a ''most probable''
electron near the ion. This most probable distance can be
estimated from the maximum of the Debye distribution $P(r)=4\pi
r^{2}n_{e}(r)/n_{e}$ $\equiv 4\pi r^{2}e^{-\beta U_{ei}(r)}$ to
give $r/r_{0}\sim \delta $ for strong coupling. The correlation
time is then approximately the time to accelerate this electron to
the position of the ion, $t_{c}\sim \left( r_{0}\delta
/\sqrt{\beta m} \right) \sqrt{1/Z\Gamma }$. The first factor
$r_{0}\delta /\sqrt{\beta m}$ is the time for an electron with the
thermal velocity to cross a sphere of the size of the thermal de
Broglie diameter. The second factor gives the dominant $Z$
dependence $t_{c}\sim \sqrt{1/Z}$. As the electron continues past
the ion its acceleration changes sign as it is attracted to the
ion with a field opposite that of the original field. This is the
source of the dominant anticorrelation.

The self-diffusion coefficient, stopping power, friction
coefficient, and width of spectral lines are proportional to the
integral of $C(t)$. At $Z=0$ this is the field autocorrelation
function at a neutral point and is independent of $Z$, depending
only on the electron-electron coupling. As $Z$ increases the
electron distribution becomes nonuniform about the ion and $ C(0)$
increases. At the same time $t_{c}$ decreases. For small $Z$ these
two effects become negligible, as seen in Fig. 6 for the case of
$\sigma =0.5Z$ and $Z<5$. As $Z$ increases, or more precisely as
$\sigma $ increases, the domain of anticorrelation appears and
begins to dominate the decrease in the integral of $C(t)$. These
basic mechanisms are captured by the meanfield description based
on the Vlasov equation for the one particle electron distribution.
The relevant correlations contained in this description are those
of the equilibrium electron distribution about the ion, also
described well by the stationary solution to the Vlasov equation.

This analysis provides a new picture for the puzzling decrease of
stopping power with increasing $Z$, relative to the Born
approximation. The stopping power is proportional to $Z^{2}$ times
the time integral of $C(t)$ which is essentially the total cross
section for all the electrons and the ion. The decrease towards a
$Z^{3/2}$ dependence at strong coupling observed earlier is seen
to be due to the effects just described. However, the precise
dependence on $Z$ may be more complicated as the coupling
increases.

Similarly, these same results provide clear evidence for the
effects of electron-electron and electron-ion correlations on the
shape of spectral lines \cite{Talin2,Talin3}. Similar experimental
puzzles regarding the $Z$ dependence of the half width \cite{3s3p}
can be clarified through combined theoretical and MD simulation as
described here.

\section{Acknowledgments}

Support for this research has been provided by the U.S. Department
of Energy Grant No. DE-FG03-98DP00218. The authors thank M.
Gigosos for helpful discussions and suggestions. J. Dufty is
grateful for the support and hospitality of the University of
Provence.

\appendix

\section{Evaluation of the field covariance}

The dimensionless electric field covariance is defined by
\begin{equation}
C(0)=\frac{r_{0}^{4}}{e^{2}}<\mathbf{E}\cdot \mathbf{E}>  \label{a.1}
\end{equation}
with
\begin{equation}
\mathbf{E}=\sum_{\alpha =1}^{N_{e}}\mathbf{e}\left(
\mathbf{r}_{\alpha }- \mathbf{r}_{0}\right) .  \label{a.2}
\end{equation}
Here $\mathbf{r}_{0}$ denotes the position of the ion. The
equilibrium average can be calculated directly in terms of the one
and two electron charge densities
\begin{equation}
C(0)=\frac{r_{0}^{4}}{e^{2}}\int d\mathbf{re}\left(
\mathbf{r}\right) \mathbf{\cdot }\left[ n_{e}(r)\mathbf{e}\left(
\mathbf{r}\right) +\int d \mathbf{r}^{\prime
}n_{e}(\mathbf{r},\mathbf{r}^{\prime })\mathbf{e}\left(
\mathbf{r}^{\prime }\right) \right]   \label{a.3}
\end{equation}
with the definitions
\begin{equation}
n_{e}(r_{1})=N\frac{\int
d\mathbf{r}_{0}d\mathbf{r}_{2}..d\mathbf{r} _{N}e^{-\beta U}}{\int
d\mathbf{r}_{0}..d\mathbf{r}_{N}e^{-\beta U}},\hspace{
0.25in}n_{e}(\mathbf{r}_{1},\mathbf{r}_{2})=N(N-1)\frac{\int
d\mathbf{r}_{0}d \mathbf{r}_{3}..d\mathbf{r}_{N}e^{-\beta U}}{\int
d\mathbf{r}_{0}..d\mathbf{r }_{N}e^{-\beta U}},  \label{a.4}
\end{equation}
where $U$ is the total kinetic energy and $T=1/k_{B}\beta $ is the
temperature.

An equivalent alternative form is obtained by writing the
covariance as
\begin{eqnarray}
C(0) &=&-\frac{r_{0}^{4}}{Ze^{3}}<\nabla _{r_{0}}U_{ie}\cdot
\mathbf{E}>= \frac{r_{0}^{4}}{\beta Ze^{3}}\frac{\int
d\mathbf{r}_{0}d\mathbf{r}_{2}..d \mathbf{r}_{N}\left( \nabla
_{r_{0}}e^{-\beta U}\right) \cdot \mathbf{E}}{
\int d\mathbf{r}_{0}..d\mathbf{r}_{N}e^{-\beta U}}  \nonumber \\
&=&\frac{r_{0}^{4}}{\beta Ze^{3}}\left\langle \nabla _{r_{0}}\cdot
\mathbf{E} \right\rangle =-\frac{r_{0}^{4}}{\beta Ze^{3}}\int
d\mathbf{r}n_{e}(r)\nabla
\mathbf{\cdot e}\left( \mathbf{r}\right)  \nonumber \\
&=&\frac{r_{0}^{4}}{\beta Ze^{3}}\int d\mathbf{r}n_{e}(r)\left(
\frac{1}{ \beta Ze}\nabla \ln n_{e}(r)\right) \mathbf{\cdot
e}\left( \mathbf{r}\right) =\frac{r_{0}^{4}}{e^{2}}\int
d\mathbf{r}n_{e}(r)\mathbf{e}_{mf}\left( \mathbf{r}\right)
\mathbf{\cdot e}\left( \mathbf{r}\right)  \label{a.5}
\end{eqnarray}
In the last equality the mean force field has been introduced by
\begin{equation}
\mathbf{e}_{mf}\left( \mathbf{r}\right) =\frac{1}{\beta Ze}\nabla \ln
n_{e}(r,t).  \label{a.6}
\end{equation}
Finally, comparison of (\ref{a.3}) and (\ref{a.5}) gives the
alternative expression for this field
\begin{equation}
\mathbf{e}_{mf}\left( \mathbf{r}\right) =\frac{1}{n_{e}(r)}\left[
n_{e}(r) \mathbf{e}\left( \mathbf{r}\right) +\int
d\mathbf{r}^{\prime }n_{e}(\mathbf{r },\mathbf{r}^{\prime
})\mathbf{e}\left( \mathbf{r}^{\prime }\right) \right] .
\label{a.7}
\end{equation}

\section{Kinetic equation for correlation functions}

In this section the evaluation of the field autocorrelation
function by kinetic theory, and the basis for the approximation
(\ref{2.16}), are briefly described. First, the correlation
function is formally rewritten as

\begin{eqnarray}
C(t) &=&\frac{r_{0}^{4}}{e^{2}}<\mathbf{E}\left( t\right)
\mathbf{\cdot E}>= \frac{r_{0}^{4}}{e^{2}}\int
d\mathbf{r}_{1}d\mathbf{v}_{1}\mathbf{..}d
\mathbf{r}_{N}d\mathbf{v}_{N}\mathbf{E\cdot E}\left( -t\right)
\rho _{e}
\nonumber \\
&=&\frac{r_{0}^{4}}{e^{2}}\int
d\mathbf{r}_{1}d\mathbf{v}_{1}\mathbf{e} \left(
\mathbf{r}_{1}\right) N\mathbf{\cdot }\int
d\mathbf{r}_{2}d\mathbf{v}
_{2}\mathbf{..}d\mathbf{r}_{N}d\mathbf{v}_{N}\mathbf{E}\left(
-t\right) \rho
_{e}  \nonumber \\
&=&\frac{r_{0}^{4}}{e^{2}}\int
d\mathbf{r}_{1}d\mathbf{v}_{1}\mathbf{e} \left(
\mathbf{r}_{1}\right) \cdot \mathbf{\psi
}(\mathbf{r}_{1},\mathbf{v} _{1};t)  \label{b.1}
\end{eqnarray}
where $\rho _{e}$ is the equilibrium Gibbs ensemble and
$\mathbf{e}\left( \mathbf{r}_{\alpha }\right) $ is the single
particle field of (\ref{2.16}). The integrations over degrees of
freedom $2..N$ in the second equality define a reduced function
$\mathbf{\psi }(\mathbf{r}_{1},\mathbf{v}_{1};t)$ which is the
first member of a set of such functions
\begin{equation}
\mathbf{\psi
}^{(s)}(\mathbf{r}_{1},\mathbf{v}_{1};..\mathbf{r}_{s},\mathbf{v
}_{s};t)=N^{s}\int
d\mathbf{r}_{s+1}d\mathbf{v}_{s+1}\mathbf{..}d\mathbf{r}
_{N}d\mathbf{v}_{N}\mathbf{E}\left( -t\right) \rho _{e}.
\label{b.2}
\end{equation}
It is straightforward to verify that these functions satisfy the
BBGKY hierarchy, whose first equation is formally the same as
(\ref{2.4})
\begin{eqnarray}
&&\left( \partial _{t}+\mathbf{v\cdot \nabla
}_{\mathbf{r}}-m_{e}^{-1}\left( \mathbf{\nabla
}_{\mathbf{r}}\left( V_{ei}(\mathbf{r})+V_{eb}(\mathbf{r} )\right)
\right) \cdot \mathbf{\nabla }_{\mathbf{v}}\right) \mathbf{\psi }(
\mathbf{r}_{1},\mathbf{v}_{1};t)  \nonumber \\
&=&m_{e}^{-1}\int d\mathbf{r}_{2}d\mathbf{v}_{2}\left(
\mathbf{\nabla }_{ \mathbf{r}}V_{ee}(\mathbf{r-r}_{2})\right)
\cdot \mathbf{\nabla }_{\mathbf{v} }\mathbf{\psi
}^{(2)}(\mathbf{r},\mathbf{v};\mathbf{r}_{2},\mathbf{v}_{2};t).
\label{b.3}
\end{eqnarray}
However, in contrast to the distribution functions in (\ref{2.4})
the functional relationship of $\mathbf{\psi }^{(2)}$ to
$\mathbf{\psi }$ is linear. To see this, consider first the
initial conditions which are found to be
\begin{eqnarray}
\mathbf{\psi }(\mathbf{r}_{1},\mathbf{v}_{1};t
&=&0)=f_{e}(\mathbf{r}_{1}, \mathbf{v}_{1})\mathbf{e}\left(
\mathbf{r}_{1}\right) +\int d\mathbf{r}_{2}d
\mathbf{v}_{2}f_{e}^{(2)}(\mathbf{r}_{1},\mathbf{v}_{1};\mathbf{r}_{2},
\mathbf{v}_{2})\mathbf{e}\left( \mathbf{r}_{2}\right)   \nonumber \\
&=&f_{e}(\mathbf{r}_{1},\mathbf{v}_{1})\mathbf{e}_{mf}(\mathbf{r}_{1})
\label{b.4}
\end{eqnarray}
\begin{eqnarray*}
\mathbf{\psi
}^{(2)}(\mathbf{r}_{1},\mathbf{v}_{1};\mathbf{r}_{2},\mathbf{v}
_{2};t
&=&0)=f_{e}^{(2)}(\mathbf{r}_{1},\mathbf{v}_{1};\mathbf{r}_{2},
\mathbf{v}_{2})\left( \mathbf{e}\left( \mathbf{r}_{1}\right)
+\mathbf{e}
\left( \mathbf{r}_{2}\right) \right)  \\
&&+n\int
d\mathbf{r}_{3}d\mathbf{v}_{3}f_{e}^{(3)}(\mathbf{r}_{1},\mathbf{v}
_{1};\mathbf{r}_{2},\mathbf{v}_{2};\mathbf{r}_{3},\mathbf{v}_{3})\mathbf{e}
\left( \mathbf{r}_{3}\right)
\end{eqnarray*}
\begin{eqnarray}
&=&f_{e}^{(2)}(\mathbf{r}_{1},\mathbf{v}_{1};\mathbf{r}_{2},\mathbf{v}
_{2})\left( \mathbf{e}_{m}\left( \mathbf{r}_{1}\right)
+\mathbf{e}_{m}\left(
\mathbf{r}_{2}\right) \right)   \nonumber \\
&&+n\int
d\mathbf{r}_{3}d\mathbf{v}_{3}h_{e}^{(3)}(\mathbf{r}_{1},\mathbf{v}
_{1};\mathbf{r}_{2},\mathbf{v}_{2};\mathbf{r}_{3},\mathbf{v}_{3})\mathbf{e}
\left( \mathbf{r}_{3}\right)   \label{b.5}
\end{eqnarray}
Here, $f_{e}^{(s)}$ are the equilibrium $s-$particle reduced
distribution functions associated with the Gibbs ensemble and
$h_{e}^{(3)}$ is the equilibrium correlation function for three
electrons in the presence of the ion
\begin{eqnarray}
h_{e}^{(3)}(\mathbf{r}_{1},\mathbf{v}_{1};\mathbf{r}_{2},\mathbf{v}_{2};
\mathbf{r}_{3},\mathbf{v}_{3};t)
&=&f_{e}^{(3)}(\mathbf{r}_{1},\mathbf{v}
_{1};\mathbf{r}_{2},\mathbf{v}_{2};\mathbf{r}_{3},\mathbf{v}_{3})
\nonumber
\\
&&-f_{e}^{(2)}(\mathbf{r}_{1},\mathbf{v}_{1};\mathbf{r}_{2},\mathbf{v}_{2})
\frac{1}{f_{e}(\mathbf{r}_{1},\mathbf{v}_{1})}f_{e}^{(2)}(\mathbf{r}_{1},
\mathbf{v}_{1};\mathbf{r}_{3},\mathbf{v}_{3})  \nonumber \\
&&-f_{e}^{(2)}(\mathbf{r}_{1},\mathbf{v}_{1};\mathbf{r}_{2},\mathbf{v}_{2})
\frac{1}{f_{e}(\mathbf{r}_{2},\mathbf{v}_{2})}f_{e}^{(2)}(\mathbf{r}_{2},
\mathbf{v}_{2};\mathbf{r}_{3},\mathbf{v}_{3})  \label{b.6}
\end{eqnarray}
The linear functional relationship between $\mathbf{\psi }^{(2)}$
to $ \mathbf{\psi }$ at $t=0$ is now evident
\begin{eqnarray}
\mathbf{\psi
}^{(2)}(\mathbf{r}_{1},\mathbf{v}_{1};\mathbf{r}_{2},\mathbf{v}
_{2};0)
&=&f_{e}^{(2)}(\mathbf{r}_{1},\mathbf{v}_{1};\mathbf{r}_{2},\mathbf{v
}_{2};t)\left(
\frac{\mathbf{\psi}(\mathbf{r}_{1},\mathbf{v}_{1};0)}{f_{e}(
\mathbf{r}_{1},\mathbf{v}_{1})}+\frac{\mathbf{\psi
}(\mathbf{r}_{2},\mathbf{v
}_{2};0)}{f_{e}(\mathbf{r}_{2},\mathbf{v}_{2})}\right)   \nonumber \\
&&+3\text{ electron correlations.}  \label{b.7}
\end{eqnarray}

Recognizing this linear relationship, the basic approximation for
weak coupling among the electrons is to neglect all of their
correlations at all times, i.e. extend (\ref{b.7}) to
\begin{equation}
\mathbf{\psi
}^{(2)}(\mathbf{r}_{1},\mathbf{v}_{1};\mathbf{r}_{2},\mathbf{v}
_{2};t)\rightarrow
f_{e}(\mathbf{r}_{2},\mathbf{v}_{2};t)\mathbf{\psi }(
\mathbf{r}_{1},\mathbf{v}_{1};t)+f_{e}(\mathbf{r}_{1},\mathbf{v}_{1};t)
\mathbf{\psi }(\mathbf{r}_{2},\mathbf{v}_{2};t).  \label{b.8}
\end{equation}
Use of this in the first hierarchy equation (\ref{b.3}) gives
directly the kinetic equation (\ref{2.16}) discussed in the text
\begin{equation}
\left( \partial _{t}+\mathcal{L}\right) \mathbf{\psi
}(\mathbf{r},\mathbf{v} ;t)=-\beta
f_{e}(\mathbf{r},\mathbf{v})\mathbf{v}\cdot \mathbf{\nabla }_{
\mathbf{r}}\int d\mathbf{r}_{2}V_{ee}(\mathbf{r-r}_{2})\int
d\mathbf{v}_{2} \mathbf{\psi }(\mathbf{r}_{2},\mathbf{v}_{2},t).
\label{b.9}
\end{equation}
\begin{equation}
\mathcal{L}=\mathbf{v\cdot \nabla }_{\mathbf{r}}-m_{e}^{-1}\left(
\mathbf{ \nabla }_{\mathbf{r}}\left( U_{ei}(\mathbf{r})\right)
\right) \cdot \mathbf{ \nabla }_{\mathbf{v}}.  \label{b.10}
\end{equation}

\section{Solution to kinetic equation}

The operator $\mathcal{L}$ in (\ref{b.3}) is the generator for
single electron dynamics in the effective potential due to the ion
$U_{ei}(\mathbf{r} )$. The solution to the equation can be
obtained in terms of this single electron dynamics by direct
integration
\begin{equation}
\mathbf{\psi
}(\mathbf{r},\mathbf{v};t)=e^{-\mathcal{L}t}\mathbf{\psi }(
\mathbf{r},\mathbf{v};0)-\int_{0}^{t}d\tau e^{-\mathcal{L}\left(
t-\tau \right) }f_{e}(\mathbf{r},\mathbf{v})\beta \mathbf{v}\cdot
\mathbf{\nabla }_{ \mathbf{r}}\int
d\mathbf{r}_{2}V_{ee}(\mathbf{r-r}_{2})\mathbf{I}(\mathbf{r}
_{2},\tau )  \label{c.1}
\end{equation}
\begin{equation}
\mathbf{I}(\mathbf{r},t)=\int d\mathbf{v\psi }(\mathbf{r},\mathbf{v},t)
\label{c.2}
\end{equation}
The initial condition is given by (\ref{b.4}). An equation for
$\mathbf{I}( \mathbf{r},t)$ follows from substitution of
(\ref{c.1}) into (\ref{c.2})
\begin{equation}
\mathbf{I}(\mathbf{r};t)=\int
d\mathbf{v}e^{-\mathcal{L}t}f_{e}(\mathbf{r},
\mathbf{v})\mathbf{e}_{mf}(\mathbf{r})-\int_{0}^{t}d\tau \int
d\mathbf{v}e^{- \mathcal{L}\left( t-\tau \right)
}f_{e}(\mathbf{r},\mathbf{v})\beta \mathbf{v }\cdot \mathbf{\nabla
}_{\mathbf{r}}\int d\mathbf{r}_{2}V_{ee}(\mathbf{r-r}
_{2})\mathbf{I}(\mathbf{r}_{2},\tau ).  \label{c.3}
\end{equation}
This is an integral equation for $\mathbf{I}(\mathbf{r};t)$ which
can be written
\begin{equation}
\int_{0}^{t}d\tau \int d\mathbf{r}_{2}\epsilon \left(
\mathbf{r},t;\mathbf{r} _{2},\tau \right)
\mathbf{I}(\mathbf{r}_{2};\tau )=\int d\mathbf{v}e^{-
\mathcal{L}t}f_{e}(\mathbf{r},\mathbf{v})\mathbf{e}_{mf}(\mathbf{r}).
\label{c.4}
\end{equation}
The dielectric function $\epsilon \left( \mathbf{r},t-\tau
;\mathbf{r} _{2}\right) $ is defined by
\begin{equation}
\epsilon \left( \mathbf{r},t-\tau ;\mathbf{r}^{\prime }\right)
=\delta \left( t-\tau \right) \delta \left(
\mathbf{r}-\mathbf{r}^{\prime }\right) +\int d\mathbf{r}^{\prime
\prime }\pi \left( \mathbf{r},t-\tau ;\mathbf{r} ^{\prime \prime
}\right) V_{ee}(\mathbf{r}^{\prime \prime }\mathbf{-r} ^{\prime
}),  \label{c.5}
\end{equation}
where the polarization function is
\begin{equation}
\pi \left( \mathbf{r},t;\mathbf{r}^{\prime }\right) =\int
d\mathbf{v}e^{- \mathcal{L}t}f_{e}(\mathbf{r},\mathbf{v})\beta
\mathbf{v}\cdot \mathbf{\nabla }_{\mathbf{r}}\delta \left(
\mathbf{r}-\mathbf{r}^{\prime }\right) . \label{c.6}
\end{equation}

With these results the correlation function from (\ref{b.1})
becomes
\begin{eqnarray}
C(t) &=&\frac{r_{0}^{4}}{e^{2}}\int d\mathbf{r}d\mathbf{ve}\left(
\mathbf{r} \right) \cdot \mathbf{\psi
}(\mathbf{r},\mathbf{v};t)=\frac{r_{0}^{4}}{e^{2}} \int
d\mathbf{re}\left( \mathbf{r}\right) \cdot
\mathbf{I}(\mathbf{r};t)
\nonumber \\
&=&\frac{r_{0}^{4}}{e^{2}}\int_{0}^{t}d\tau \int
d\mathbf{re}\left( \mathbf{r }\right) \cdot \int
d\mathbf{r}_{2}\epsilon ^{-1}\left( \mathbf{r},\tau ;
\mathbf{r}_{2},\right) \int d\mathbf{v}e^{-\mathcal{L}\left(
t-\tau \right)
}f_{e}(\mathbf{r}_{2},\mathbf{v})\mathbf{e}_{mf}(\mathbf{r}_{2})
\nonumber
\\
&=&\frac{r_{0}^{4}}{e^{2}}\int_{0}^{t}d\tau \int
d\mathbf{r}d\mathbf{\mathbf{v}e}_{s}\left( \mathbf{r},\tau \right)
\cdot e^{-\mathcal{L}\left( t-\tau \right)
}f_{e}(\mathbf{r},\mathbf{v})\mathbf{e}_{mf}(\mathbf{r}).
\label{c.7}
\end{eqnarray}
In the last equality the screened field $\mathbf{e}_{s}\left(
\mathbf{r} ,t\right) $ has been introduced
\begin{equation}
\mathbf{e}_{s}\left( \mathbf{r},t\right) \equiv \int
d\mathbf{r}^{\prime } \mathbf{e}\left( \mathbf{r}^{\prime }\right)
\epsilon ^{-1}\left( \mathbf{r} ^{\prime },t;\mathbf{r}\right)
\label{c.8}
\end{equation}
Finally, using the stationarity of $f_{e}(\mathbf{r},\mathbf{v})$
under the dynamics generated by $\mathcal{L}$ gives
\begin{eqnarray}
C(t) &=&\frac{r_{0}^{4}}{e^{2}}\int_{0}^{t}d\tau \int
d\mathbf{r}d\mathbf{
\mathbf{v}}f_{e}\mathbf{(\mathbf{r},\mathbf{v})e}_{s}\left(
\mathbf{r},\tau \right) \cdot e^{-\mathcal{L}\left( t-\tau \right)
}\mathbf{e}_{mf}(\mathbf{r
})  \nonumber \\
&\equiv &\frac{r_{0}^{4}}{e^{2}}\int_{0}^{t}d\tau \int
d\mathbf{r}d\mathbf{
\mathbf{v}}f_{e}\mathbf{(\mathbf{r},\mathbf{v})e}_{s}\left(
\mathbf{r},\tau \right) \cdot \mathbf{e}_{mf}(\mathbf{r}\left(
t-\tau \right) ).  \label{c.9}
\end{eqnarray}
This is the form used in the text, Eq. (\ref{2.19}).

\end{document}